\documentclass[twocolumn,aps,prl,amsmath,amssymb]{revtex4-1}
\usepackage{physics}
\usepackage{graphicx}
\usepackage{mwe}
\usepackage{bm}
\usepackage{epstopdf}
\usepackage{subfig}
\usepackage{epsfig}
\usepackage{lipsum}
\usepackage{epstopdf}
\usepackage{epsfig}
\usepackage{dsfont}
\usepackage{amssymb}

\usepackage{braket}

\usepackage{leftidx}
\begin{document}

\title{Quantum-state texture and gate identification}
%\title{Quantum texture and the identification of unknown circuit layers}
%\title{Quantum texture and the identification of CNOT gates in an unknown circuit layer}

\author{Fernando Parisio}
\email[]{fernando.parisio@ufpe.br}
\affiliation{Departamento de
		F\'{\i}sica, Centro de Ci\^encias Exatas e da Natureza, Universidade Federal de Pernambuco, Recife, Pernambuco
		50670-901 Brazil}

\begin{abstract}
We introduce and explore the notion of texture of an arbitrary quantum state, in a selected basis. In the first part of this letter we develop a resource theory and show that state texture is adequately described by an easily computable monotone, which is also directly measurable. 
%We give a simple example of an ideal paramagnet, relating the texture of a class of non-equilibrium states with the magnetization of the corresponding thermalized state. 
It is shown that textures are useful in the characterization of unknown quantum gates in universal circuit layers. By using randomized input states and recording the textures of the output qubits we are able to fully characterize the circuit layer, whenever it contains at least one CNOT gate. This can be done without the need of tomographic protocols and the use of ancillary systems.
\end{abstract}

\maketitle
%\section{Introduction}
%

{\it Introduction}. The notorious quote attributed to C. Caves, ``Hilbert space is a big place'' (more accurately, ``Hilbert space is gratuitously big''\cite{caves}), refers to the exponential growth of the dimension of quantum state space with the number of subsystems. So, even describing the state of a relatively modest system can quickly become a prohibitive task. In this context, the emergence of a reduced number of quantities that provide relevant, albeit partial, information about a quantum system is not surprising.
Notable examples of such figures of merit include purity $P$ \cite{purity, purity2} and the $\ell_1$-norm of coherence $C_1$ \cite{l1}.
These simple quantities not only illuminate foundational aspects of quantum theory but have also been demonstrated to be essential resources across various practical applications. 

It should be noted that these examples correspond to sums involving the absolute values of the density matrix entries, 
$\varrho_{ij}$ ($P=\sum_{i,j}|\varrho_{ij}|^2$, $C_1=\sum_{i\ne j}|\varrho_{ij}|$). Other noteworthy instances include the $\ell_q$-norms of coherence ($C_q=\sum_{i\ne j}|\varrho_{ij}|^q$) and various measures of purity and localization \cite{purity-loc}. These figures of merit are insensitive to the local phases attached to the coherences, and, to use the term coined in \cite{imag0}, do not capture the ``imaginarity'' of quantum states. This aspect is crucial in quantum state discrimination \cite{imag}, for instance, and is ultimately an inextricable feature of quantum coherence and quantum mechanics itself \cite{nature}. Coherence \cite{review,girolami,yuan,adesso,rana,bu,xu}, as the potential for coexisting circumstances that would be mutually exclusive in classical contexts, is too complex to be fully described by a single figure of merit.

It is also an essential component for emerging technologies, ranging from quantum computation \cite{hil,pati,ma,matera,pan,pezze,felix,nostro} to metrology \cite{metro,metro2}. 
Almost invariably, these technologies involve the controlled manipulation of quantum states, analogous to the operation of a quantum circuit. Consequently, the characterization of these structures is of paramount importance \cite{poyatos, nielsen, gus,gates,duan, bench,preskill}. This task is resource-intensive, as evidenced by the fact that the number of observables required to be measured at the output state scales as $O(D^2)$, where $D$ is the dimension of the system \cite{gus}. Depending on the protocol, the necessary number of distinct input states also grows with $D$.

One of the objectives of this letter is to introduce a simple quantity, akin to $C_1$, but sensitive to imaginarity, from a formal resource-theoretical perspective. The second part of this letter focuses on employing the introduced concept (quantum-state texture) for a novel approach to quantum gate identification. We utilize random inputs and measurements of state texture across two bases and consider universal circuit layers. When the circuit is initialized with identical input qubits, the randomness appearing in the statistics of distinct runs, one is able to fully characterize the circuit layer.

\begin{figure}[h]
  \includegraphics[height=2.7cm]{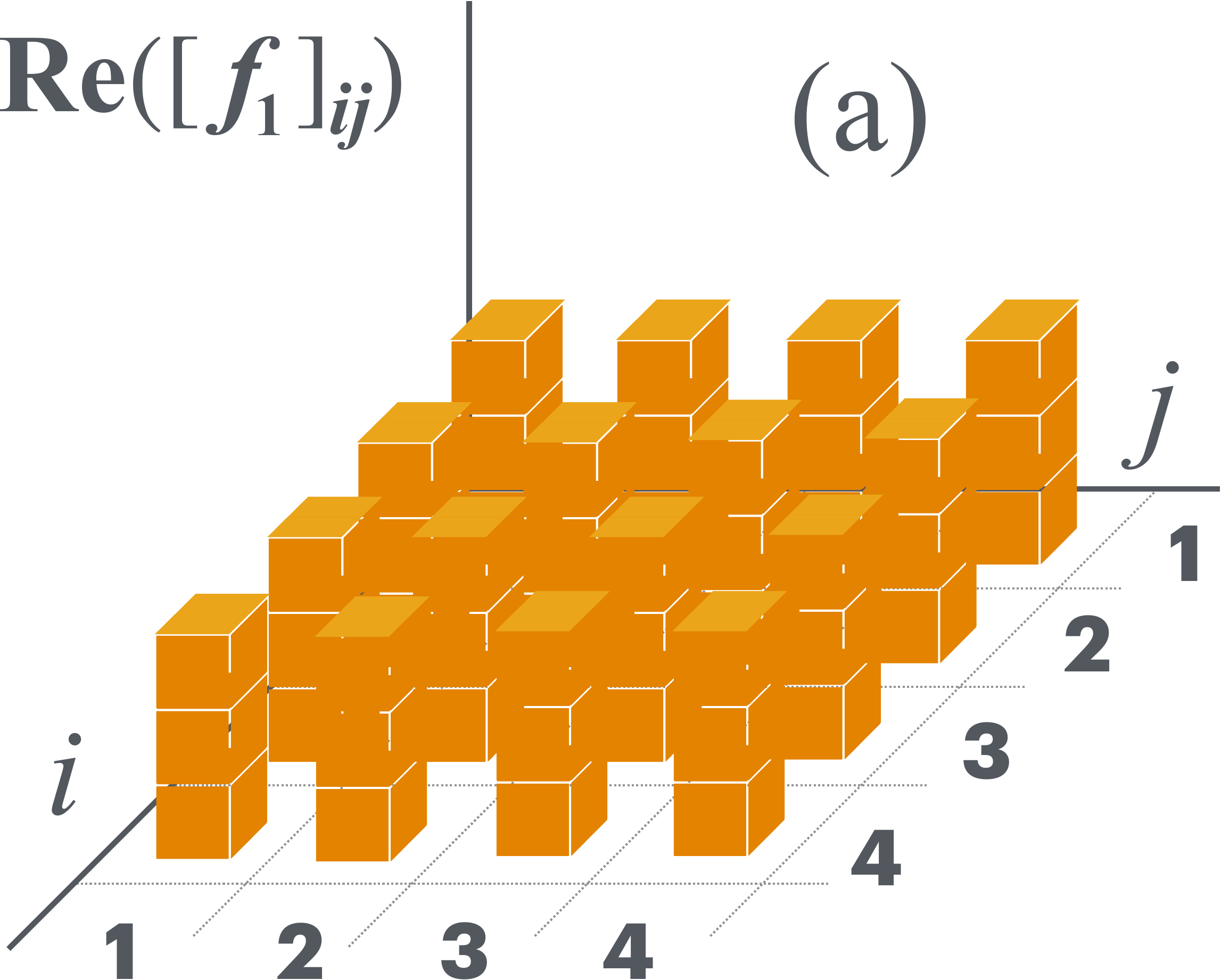}
    \includegraphics[height=2.7cm]{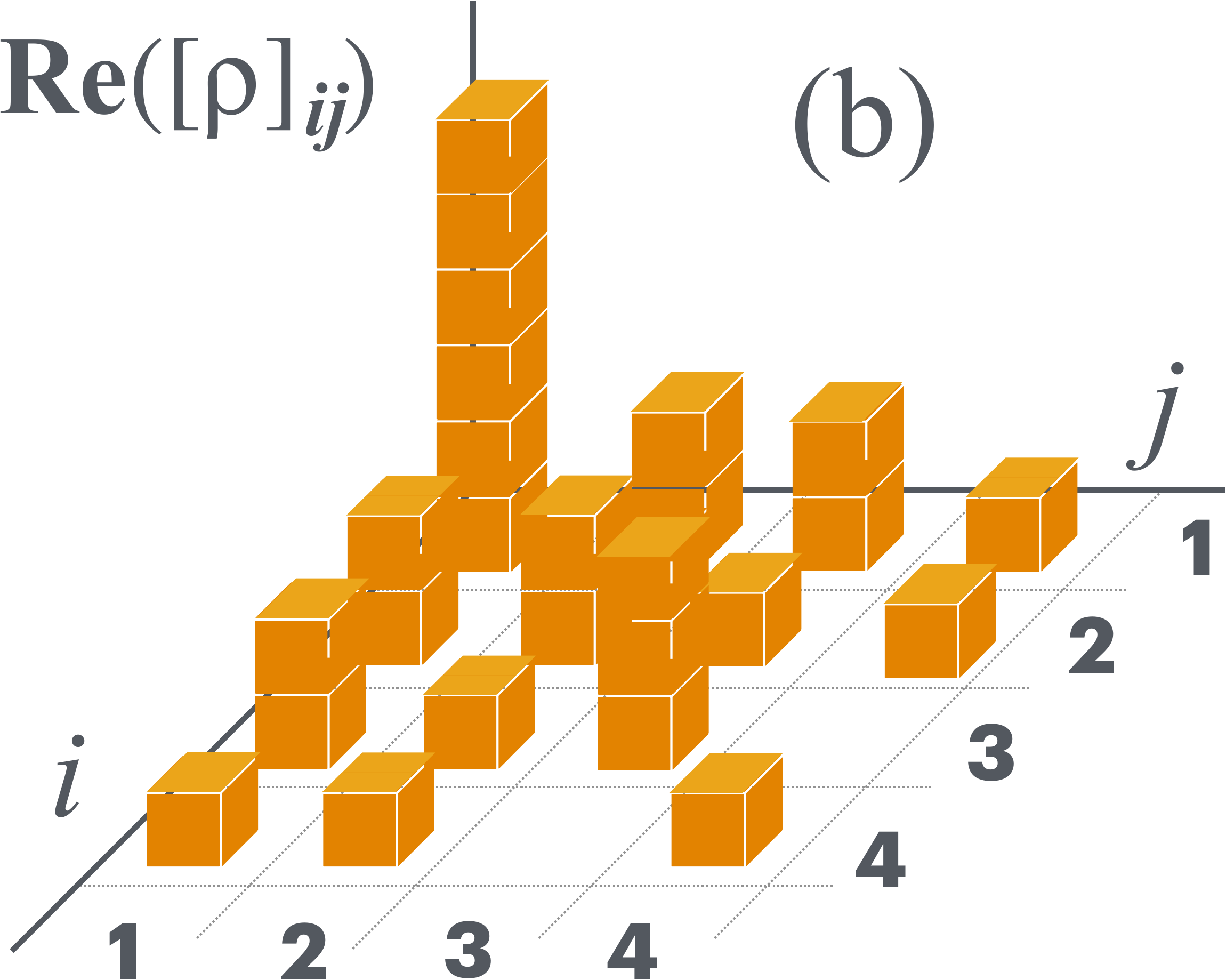}
  \caption{Pictorial bar charts displaying the real part of density-matrix entries with $D=4$: (a) the textureless state $f_1$ and (b) an arbitrary state $\rho$. Note that ${\rm Im}([f_1]_{ij})=0$ for all pairs $i$, $j$.}
  \label{fig1}
\end{figure}

%\section{State Texture}
%
{\it State Texture}. Given a fixed basis, hereafter denoted by $\{ |i \rangle\}$ and referred to as the computational basis, one can see a density operator $\varrho$ as a table, with each cell corresponding to a matrix entry. A common way to visualize this table is to provide a tridimensional plot where the third dimension corresponds to the real part of each $\varrho_{ij}$ (and an analogous plot for the imaginary part), see Fig. \ref{fig1}. The landscape formed by these bar charts, in general, display some irregularity, or texture. In this regard, the simplest possible table is the one for which all cells are filled with the same number. The only existing quantum state which gives rise to such a table is $f_1\equiv |f_1\rangle \langle f_1| \in {\cal B(H)}$, with $ |f_1\rangle \equiv \frac{1}{\sqrt{D}}\sum_{i=1}^D|i\rangle$, where $D$ is the dimension of the Hilbert space ${\cal H}$. We refer to this state as textureless and it alone corresponds to the zero-resource set of the theory to be developed. 

The associated free operations, corresponding to completely positive and trace preserving maps $\Lambda$, are effected by Kraus operators, $\Lambda(\varrho)=\sum_nK_n\varrho K^{\dagger}_n$, for which $\sum_n K_n^{\dagger}K_n=\mathds{1}$. These free maps must not create texture and, thus, $f_1$ should be a fixed point, $\Lambda(f_1)=f_1$. Note that, since $f_1$ is a pure state and, thus, an extremal point of ${\cal B(H)}$ we must have $K_n|f_1\rangle \propto |f_1\rangle$ for all $n$.

In addition, given an arbitrary texture measure, ${\cal T}$, it must be such that  {\bf (i)} ${\cal T}(f_1)=0$ and  {\bf (ii)} ${\cal T}(\varrho)\ge {\cal T}(\Lambda(\varrho))$ for all $\varrho \in {\cal B(H)}$.
Finally, we assume that {\bf (iii)} state texture must not increase under mixing: ${\cal T}(\sum p_i \varrho_i) \le \sum p_i {\cal T}(\varrho_i)$. 

Since we have a single ket as the zero-resource set, it is reasonable to expect that an appropriate function of $\langle f_1|\varrho|f_1 \rangle$, should be a good candidate for a quantifier of the texture of $\varrho$. 
%As we will see, this is indeed the case.  
This quantity has a direct geometric interpretation, the overlap between the state $\varrho$ and the textureless state. Curiously, it also embodies a simple algebraic interpretation. Given an arbitrary state $\varrho$, we have 
\begin{equation}
D \langle f_1|\varrho|f_1 \rangle= \sum_{i, j=1}^D\langle i |\varrho | j \rangle= \sum_{i, j=1}^D\varrho_{ij} \equiv \Sigma(\varrho),
\end{equation}
corresponding to the sum of all matrix elements of $\varrho$ in the chosen basis, sometimes, referred to as the grand sum of $\varrho$. Since for any state the contribution of the diagonal entries is fixed [${\rm Tr}(\varrho)=1$], the grand sum gives relevant information about the coherences. 

As important as its computable form is the fact that $\Sigma(\varrho)$ is a directly measurable quantity. This follows from the identity $\Sigma(\varrho)/D={\rm Tr}(\varrho |f_1\rangle \langle f_1|)=P_{f_1}$. Therefore $\Sigma(\varrho)/D$ is the probability of obtaining $|f_1\rangle$ in a projective measurement of any observable which has $|f_1\rangle$ as an eigenvector. This greatly facilitates the experimental determination of $\Sigma(\varrho)$, excluding, for instance, the need of characterizing $\varrho$ tomographically.
Note also that $0\le \Sigma(\varrho)\le D$, the maximum value being attained for the zero-resource state. Based on this observation we define a texture quantifier, the state rugosity, simply as
\begin{equation}
\mathfrak{R}(\varrho)=-\ln \left(\frac{\Sigma(\varrho)}{D}\right)=-\ln\langle f_1|\varrho|f_1 \rangle,
\end{equation}
with $\mathfrak{R} \in [0, \infty)$. Note that $\mathfrak{R}(\varrho)=0$ if and only if $\varrho=f_1$, which is stronger than ${\bf (i)}$, ensuring that $\mathfrak{R}$ is faithful. 
We observe that, for an arbitrary qubit state $\varrho=\frac{1}{2}(\mathds{1}+x\sigma_x+y\sigma_y+z\sigma_z)$, we have $\mathfrak{R}(\varrho)=\ln(\frac{2}{1+x})$. 
Interestingly, the same state have imaginarity $M(\varrho)=2|y|$, as given by the measure introduced in \cite{imag0}, illustrating the complementary nature of the two quantities.

In the supplemental  material (SM) we prove that $\mathfrak{R}(\varrho)$ is non-increasing under general free operations and for an arbitrary $\varrho$, thus, satisfying {\bf (ii)}. 
Notice that $\mathfrak{R}$ remains invariant under mixing, saturating, but still fulfilling {\bf (iii)}. This follows immediately from 
$\Sigma\left(\sum_k p_k \varrho^{(k)}\right)=\sum_{i,j,k} p_k \varrho^{(k)}_{ij}=\sum_k p_k \Sigma( \varrho^{(k)})$,
or
\begin{equation*}
\mathfrak{R} \left(\sum_k p_k \varrho^{(k)}\right)=\sum_k p_k \mathfrak{R}( \varrho^{(k)}).
\end{equation*} 

As for the states that maximize $\mathfrak{R}$, they are defined in the orthogonal support of $f_1$, corresponding to states with a vanishing grand sum. For arbitrary $D$, this is exactly the case of all but one Fourier states:
\begin{equation}
\label{fourier}
|f_k\rangle=\frac{1}{\sqrt{D}}\sum_{j=1}^D \omega_D^{(k-1)(j-1)}|j\rangle,
\end{equation}
where $\omega_D=e^{2\pi i/D}$ and $k=1, 2 \dots, D$. For $k=1$ we get $|f_1\rangle$, the zero-texture state. For states with $k>1$ we have $\mathfrak{R}(f_k)\rightarrow \infty$, where $f_k \equiv |f_k\rangle \langle f_k|$, with $\Sigma\propto |\sum_j\omega_D^{j-1}|^2=0$. Note that any linear combination of the kets $|f_k \rangle$, with $k>1$ also maximizes $\mathfrak{R}$. 

However, saturating a particular quantifier is not sufficient to conclude that a state contains the maximum amount of resource. In the SM we demonstrate that any state, either pure or mixed, can be produced from the Fourier states $|f_k\rangle$, with $k>1$, via free operations. Therefore, these Fourier states are also maximal in a formal resource-theoretical sense. This completes the essentials of the quantum-state texture resource theory.

Before addressing a specific system, we show that $\mathfrak{R}$ is additive, a useful property that is seldom valid for quantum measures
(one exception is the logarithmic negativity \cite{vidal, plenio}).
Consider two arbitrary states $\rho$ and $\sigma$ associated with Hilbert spaces of dimensions $D_1$ and $D_2$, respectively. One can write $\Sigma(\rho \otimes \sigma)=\sum_{i_1,j_1=1}^{D_1} \sum_{i_2,j_2=1}^{D_2} \langle i_1 \,i_2| \rho \otimes \sigma |j_1 \, j_2 \rangle=\sum_{i_1,j_1=1}^{D_1}\langle i_1| \rho  |j_1 \rangle \, \sum_{i_2,j_2=1}^{D_2} \langle i_2| \sigma | j_2 \rangle=\Sigma(\rho) \, \Sigma(\sigma)$.
Therefore, $\mathfrak{R}(\rho \otimes \sigma)=\mathfrak{R}(\rho)+\mathfrak{R}(\sigma)$, or more generally 
\begin{equation*}
\label{add}
\mathfrak{R}\left(\bigotimes_{\nu=1}^N \rho_{\nu} \right)=\sum_{\nu=1}^N\mathfrak{R}(\rho_{\nu}),
\end{equation*}
where $N$ is the number of subsystems. In particular, this property also implies $\mathfrak{R}(\rho^{\otimes n})=n\mathfrak{R}(\rho)$. 

In the SM we investigate quantum-state texture in a simple physical system: an ideal paramagnet in contact with a thermal bath of absolute temperature $T$. In thermodynamic equilibrium, the rugosity is a constant, independent of $T$, $\mathfrak{R}=\ln D$. In this context, a variation in quantum texture as the temperature changes is an indicator of a non-equilibrium situation. By investigating a specific class of non-equilibrium states, with the spin chain in thermal contact with the same reservoir, we show that  the rugosity of the non-equilibrium state is completely determined by the magnetization of the corresponding system in thermal equilibrium. 
For this result to hold we necessarily have to employ uniformly distributed phases, thus requiring the so-called imaginarity of quantum states \cite{imag0,imag,nature}.

%\section{Quantum-circuit characterization}

\begin{figure}[h]
  \includegraphics[height=2.8cm]{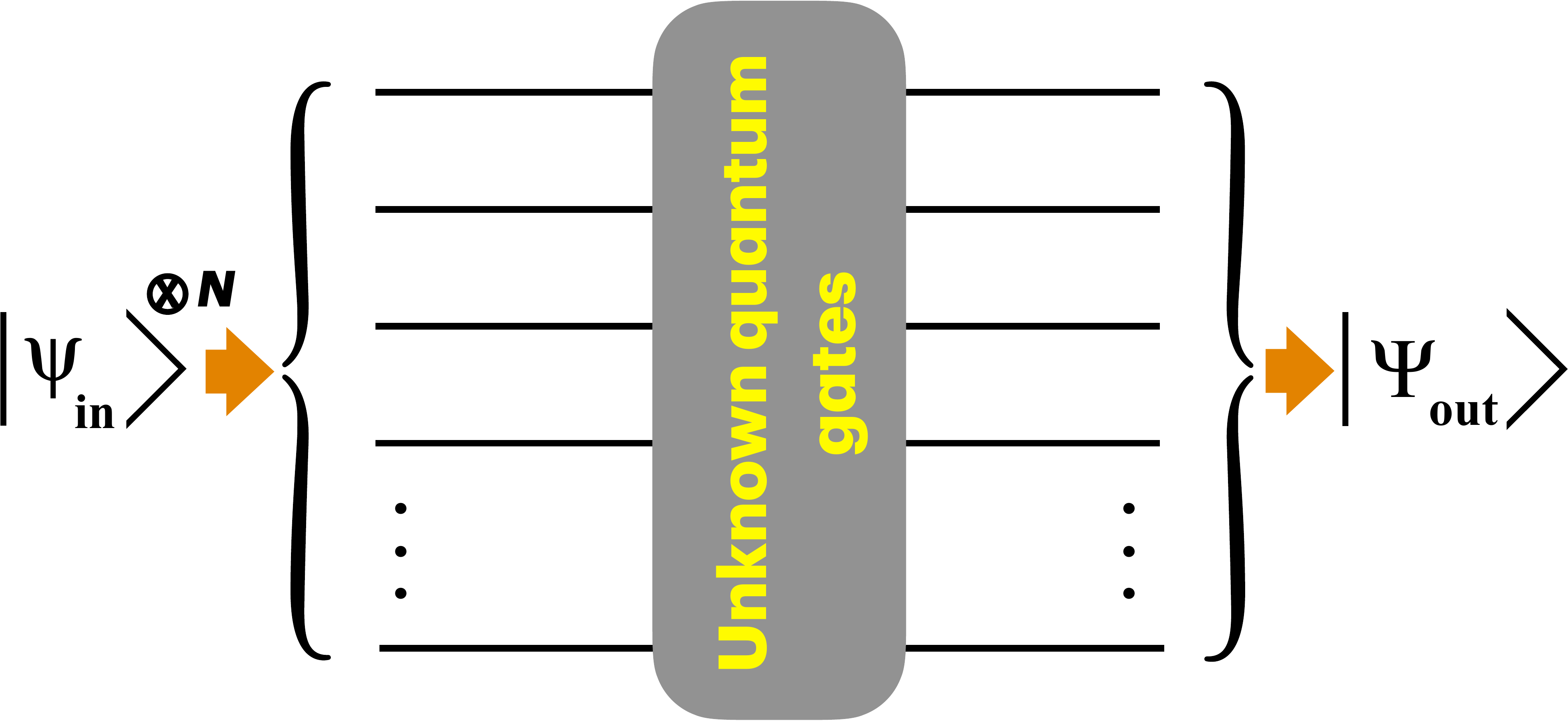}
  \caption{In the circuit layer any qubit is either acted upon by a single gate or by no gate.}
  \label{fig2}
\end{figure}

{\it Quantum-circuit characterization}.  One important practical problem in quantum technologies is the identification of processes. The characterization of unknown quantum gates and circuit layers is particularly relevant in this program \cite{gates, bench}. The problem is usually formulated as follows: providing the input and knowing the output states $\varrho_{\rm out}=U\varrho_{\rm in}U^{-1}$, one has to devise a series of input states and POVMs on $\varrho_{\rm out}$ which enable the reconstruction of the unknown unitary $U$.
As an application of the texture concept, we consider a single layer of quantum gates with an arbitrary number of input qubits (Fig. \ref{fig2}) and assume that the employed gates are arbitrary one-qubit gates plus CNOT gates. For instance, one can choose
\begin{eqnarray}
\nonumber
U_{\rm CN}&=&\begin{pmatrix}
1 & 0 & 0 & 0\\
0 & 1 & 0 & 0 \\
0 & 0 & 0 & 1 \\
0 & 0 & 1 & 0 \\
\end{pmatrix}, \;
U_{\rm H}=\frac{1}{\sqrt 2}\begin{pmatrix}
1 & 1 \\
1 & -1  \\
\end{pmatrix},\\
U_{\rm T}&=&\begin{pmatrix}
e^{-i\pi/8} & 0 \\
0 & e^{i\pi/8}  \\
\end{pmatrix},\;
U_{\rm S}=\begin{pmatrix}
1 & 0 \\
0 & i \\
\end{pmatrix},
\label{gates}
\end{eqnarray}
which form a universal set for fault-tolerant computation \cite{Ugates}. Of course we do not know the basis, denoted by $\{|\pm\rangle \}$, in which these gates assume their standard form, Eq. (\ref{gates}).
Here, by a circuit layer we mean a portion of the circuit in which each qubit is either acted upon by a single gate or by no gate (corresponding to the identity operator). More precisely, the first layer begins whenever any of the circuit qubits is first acted upon by a gate, and ends just before any of the qubits is acted upon by a gate, for a second time (and so on).  In what follows, we show that by using random-input states (for a recent reference see \cite{random}) and measuring only two non-entangled observables at the output qubits, we can find the unknown basis. 

Notice that, due to the isotropy of random qubits over the Bloch sphere and the averaging process, one can express them as $|\psi_{\rm in}\rangle=\cos(\theta/2)|+\rangle+e^{i\phi}\sin(\theta/2)|-\rangle$, even though we do not know the basis $\{|\pm\rangle \}$, with $0\le \theta \le \pi$ and $0 \le \phi \le 2\pi$, distributed according to the Haar measure. After the qubits are acted upon by the gates one determines the grand sum of each individual output state, with respect to the arbitrarily chosen basis $\{|1\rangle, |2\rangle \}$. We recall that finding the grand sum with respect to the computational basis means, in fact, to project in the basis  $\{|f_1\rangle, |f_2\rangle \}$, and vice versa. 

Note carefully: we assume that, in each run, all input qubits are the same, the randomness showing up when different runs are considered. Since one cannot simply clone a single random qubit, this poses technical difficulties, which, fortunately, can be dealt with, either by non-ideal, though efficient cloning \cite{clone1,clone2,clone3,clone4,clone4b,clone5,clone6,clone7,clone8}, or via other preparation techniques, see e. g., \cite{hamoudi}. 

In the  supplemental  material we show that, by adopting this procedure a sufficient number of runs, one gets the single-qubit average $\overline{\Sigma}_{\rm out}^{\Gamma}=1$, for an arbitrary single-qubit gate $\Gamma$. Regarding our particular example in Eq. (\ref{gates}):
$\overline{\Sigma}_{\rm out}^{\rm T}=\overline{\Sigma}_{\rm out}^{\rm S}=\overline{\Sigma}_{\rm out}^{\rm H}=1=\overline{\Sigma}_{\rm in}$,
where the averages are over the output states of a fixed circuit track. Of course, we also have
$\overline{\Sigma}_{\rm out}^{\mathds{1}}=1$ for the qubits that do not go through any gate. The randomness of the inputs wipes out any information from these average grand sums. 

The output qubits that take part in a CNOT gate do provide relevant information. This may sound counterintuitive, since the input qubits are random, after all. CNOT gates act on a two-qubit space and the key point is that the tensor product of the two qubits $|\psi_{\rm in}\rangle\otimes |\psi_{\rm in}\rangle$, individually satisfying the Haar measure, is not a Haar random ququart. The coefficients of $|+-\rangle$ and  $|-+\rangle$ are, e. g., always the same. Due to this, the average grand sums related to these qubits are not 1, in general.

Let us write $|+\rangle=\alpha|1\rangle+\beta|2\rangle$ and $|-\rangle=\beta^*|1\rangle-\alpha^*|2\rangle$, where $\alpha$ and $\beta$ $\in \mathds{C}$ are unknown constants, apart from the constraint $|\alpha|^2+|\beta|^2=1$. In the SM we show that
\begin{eqnarray}
\label{cnot1a}
\overline{\Sigma}_{\rm out}^{\mbox{\large \textbullet}}&=&1-\frac{1}{6}(\alpha^2+{\alpha^*}^2-\beta^2-{\beta^*}^2),\\
\label{cnot1b}
\overline{\Sigma}_{\rm out}^{\oplus}&=&1+\frac{1}{3}({\alpha^*}\beta+\alpha{\beta^*}),
\end{eqnarray}
where ``{\large \textbullet}'' (``$\oplus$'') refers to qubits that served as control (target) in CNOT gates. For almost all bases $\{|1\rangle, |2\rangle \}$ these last averages are different from 1, enabling the identification of qubits that went through CNOT gates. 

To complete the protocol a second observable is needed. Not surprisingly this observable is related to the grand sum in the Fourier basis $\{|f_1\rangle, |f_2\rangle \}$. In this basis, the textureless state is simply $(|f_1\rangle+ |f_2\rangle )/\sqrt{2}=|1\rangle$. Therefore, the second measurement is related to projections onto $\{|1\rangle, |2\rangle \}$. Since the role of these bases is interchangeable, one can immediately write $\alpha \rightarrow (\alpha+\beta)/\sqrt{2}$ and $\beta \rightarrow (\alpha-\beta)/\sqrt{2}$ in Eqs. (\ref{cnot1a}) and (\ref{cnot1b}), to get
\begin{eqnarray}
\label{cnot2a}
\overline{\Sigma'}_{\rm out}^{\mbox{\large \textbullet}}&=&1-\frac{1}{3}({\alpha}\beta+{\alpha^*}{\beta^*}),\\
\label{cnot2b}
\overline{\Sigma'}_{\rm out}^{\oplus}&=&1+\frac{1}{3}(|\alpha|^2-|\beta|^2).
\end{eqnarray}
We remark that, although we know that (\ref{cnot1a}) and (\ref{cnot2a}) refer to the same qubits (same tracks in the circuit), analogously for the pair (\ref{cnot1b}) and (\ref{cnot2b}) and, since the measured quantities fall within the interval $[2/3,4/3]$, the experimenter has no means to decide which pair of data corresponds to control and target. So we must consider the associations
\begin{equation}
\label{notation}
(X=\overline{\Sigma}_{\rm out}^{\mbox{\large \textbullet}},\tilde{X}=\overline{\Sigma}_{\rm out}^{\oplus}), \;\;(Y=\overline{\Sigma'}_{\rm out}^{\mbox{\large \textbullet}},\tilde{Y}=\overline{\Sigma'}_{\rm out}^{\oplus});
\end{equation}
and the other way around: $X \leftrightarrow \tilde{X}$ and $Y \leftrightarrow \tilde{Y}$, i. e, $\oplus \leftrightarrow \mbox{\large \textbullet}$.  
Importantly, we show that the inequality (SM)
\begin{equation}
\label{ineq}
\Delta_{\mbox{\large \textbullet}}+\Delta_{\oplus}\ge\frac{1}{9}
\end{equation}
holds no mater the choice of the computational basis, where $\Delta_{z}=\left[\left(\overline{\Sigma}_{\rm out}^{z}-1\right)^2+\left(\overline{\Sigma'}_{\rm out}^{z}-1\right)^2\right]$, $z=\oplus, \mbox{\large \textbullet}$. 
Therefore, the hypothetical situation where $\label{control}
\overline{\Sigma}_{\rm out}^{\mbox{\large \textbullet}}=\overline{\Sigma}_{\rm out}^{\oplus}=\overline{\Sigma'}_{\rm out}^{\mbox{\large \textbullet}}=\overline{\Sigma'}_{\rm out}^{\oplus}=1$ is a mathematical impossibility.
In addition, the situation in which three out of the four average grand sums equal 1, corresponds to a zero-measure set. Even in this case, inequality (\ref{ineq}) ensures that the distinct grand sum reaches a limiting value (either $2/3$ or $4/3$), which allows for unambiguous identification of CNOT gates. More generally, this inequality implies that in any situation, it is not possible that all four averages are arbitrarily close to 1. Consider, for instance, that $X$, $\tilde{X}$, $Y$, and $\tilde{Y}$ all differ from 1 by $\pm \delta$. In this case, we must have $|\delta|\ge 1/6$, which provides a sizable difference from the averages associated with single-gate qubits. So, one can always detect the CNOT gates.

At this point, we know how many such gates there are and the position of the involved qubits. This is already a valuable information because CNOT gates often determine the running time of the code (as is the case of ion-trap quantum computers), and impose limits to the global fidelity of the circuit \cite{reviewIon}.  They also cause an increase in the circuit complexity, which have motivated efforts to reduce their count  in specific situations \cite{downcnot, downcnot2}.

But this is not all, since the previous data suffices to decrease the possibilities for the unknown basis from infinitely many to four. 
By solving equations (\ref{cnot1a}) to (\ref{cnot2b}) in terms of the measured grand sums we find the once unknown parameters $\alpha\equiv|\alpha|e^{i\lambda}$ and $\beta\equiv|\beta|e^{i\chi}$. They are given by the expressions:
\begin{eqnarray}
\label{final}
\nonumber
|\alpha|&=&\sqrt{\frac{3}{2}\tilde{Y}-1},\;\; |\beta|=\sqrt{2-\frac{3}{2}\tilde{Y}},\\
\nonumber
\cos \lambda&=& \pm\frac{\sqrt{\tilde{Y}-X+\sqrt{(\tilde{Y}-X)^2+(\tilde{X}-Y)^2}}}{\sqrt{2\tilde{Y}-\frac{4}{3}}},\\
\nonumber
\cos \chi&=& \pm\frac{\sqrt{X-\tilde{Y}+\sqrt{(\tilde{Y}-X)^2+(\tilde{X}-Y)^2}}}{\sqrt{\frac{8}{3}-2\tilde{Y}}},
\end{eqnarray}
for $\alpha\ne 0$, $\beta \ne 0$, and where the signs can be combined independently. This result does not depend on the number of qubits. In the optimal situation where  $\alpha= 0$ or $\beta = 0$, our chosen basis already coincides with the unknown basis, apart from relabelings. In these expressions we assumed that (\ref{notation}) holds and in the SM we show that considering the possibility $\tilde{X}\leftrightarrow X$ and $\tilde{Y}\leftrightarrow Y$ and the possible phases, there remains a set of four possible bases to be tested. Once we get to this point, we show that it is trivial to single out the correct basis. This would finish the complete characterization of the circuit layer, without any state or process tomography and without the need of POVMs, and thus, of ancillary systems.

Of course, actual states and gates are imperfect. Although a detailed analysis of realistic noise models is beyond the scope of this letter, we provide some preliminary comments on white-noise admixture. Specifically, if we suppose that the input qubits have a white-noise fraction $p$ and that the CNOT gates function simply as the identity operator for a fraction $q$ of the input qubits, then, the measured grand sums would fall in the interval
$[1-(1+qp-p-q)/3,1+(1+qp-p-q)/3]$. For instance, if we have $p=0.2$ and $q=0.3$ the interval becomes $[0.813,1.187]$, which would still enable the identification of CNOT gates. In this case, one can determine their number, but the full identification of the unknown basis would require a previous characterization of the noise levels. Note that, the noise related to one-qubit gates is unimportant for the proposed protocol.

%\section{Summary}
%

{\it Summary and perspectives}. One generically links uniformity with simplicity, whereas its lack relates to some notion of intricacy. We have herein formalized this notion to establish a resource theory of state texture.
We introduced a {\it bona fide} measure that does not require numeric optimizations, is additive, non-increasing under mixing, and experimentally friendly. It was shown that evaluating the texture of output qubits emerging from an unknown circuit layer enable us to fully characterize it. 

The possibilities of future investigations are broad. Texture is clearly related to quantum coherence \cite{review} and imaginarity \cite{imag0,imag,nature} and 
a deeper understanding of the interplay among these quantities is an interesting venue of research.  Also, since coherence is relevant to quantum thermodynamics \cite{bertulio,prxgour} we intend to investigate the possible role of quantum texture (in the energy eigenbasis), in this context. The study of quantum-texture dynamics, including open systems, in similar lines as those of \cite{dynamics}, for the $\ell_1$-norm of coherence, seems to be worthwhile. It would be interesting to compare the time evolution of these two somewhat complementary features of quantum coherence. 

In a distinct front, it is believed that some fundamental biological phenomena are due to a remanent level of quantum coherence, so, a potentially fruitful line of investigation is to address the role of state texture in quantum biology, in the spirit of references \cite{huelga,fuller,groot}.

\begin{acknowledgments}
		This work received financial support from the Brazilian agencies Coordena\c{c}\~ao de Aperfei\c{c}oamento de Pessoal de N\'{\i}vel Superior (CAPES), Conselho Nacional de Desenvolvimento Cient\'{\i}fico  e Tecnol\'ogico through its program CNPq INCT-IQ (Grant 465469/2014-0), Funda\c{c}\~ao de Amparo \`a Pesquisa do Estado de S\~ao Paulo (FAPESP - Grant 2021/06535-0), and Funda\c{c}\~ao de Amparo \`a Ci\^encia e Tecnologia do Estado de Pernambuco (FACEPE - Grant BPP-0037-1.05/24).
	\end{acknowledgments}

\begin{widetext}
\begin{center}
{\large Supplemental  Material on ``Quantum-state texture and gate identification''}\\
\vspace{0.6cm}
Fernando Parisio\\
\vspace{0.1cm}
Departamento de
		F\'{\i}sica, Universidade Federal de Pernambuco, Recife, Pernambuco
		50670-901 Brazil
\end{center}		
\vspace{0.1cm}

\section{ Rugosity does not increase under general free operations}
We start with the general requirements that must be satisfied by the free maps ($\Lambda$). Given an arbitrary state $\varrho$, in the operator sum representation we have $\Lambda(\varrho)=\sum_nK_n\varrho K^{\dagger}_n$, with
\begin{equation}
\label{C1}
\sum_n K^{\dagger}_n K_n=\mathds{1}, 
\end{equation}
for any physical map and
\begin{equation}
\label{C2}
\Lambda(f_1)=\sum_nK_n f_1 K^{\dagger}_n=f_1,
\end{equation}
which must be valid for the free operations under investigation.

As we mentioned in the main text, the only possibility to fulfill condition (\ref{C2}) is $K_n|f_1 \rangle=a_n|f_1\rangle$, for all $n$, where $a_n$ is a constant. This is due to the fact that pure states are extremal points of the Hilbert-Schmidt space ${\cal B(H)}$ and as such have a single, trivial representation in terms of convex combinations. Next, we define the operators $A_n\equiv K_n-a_n|f_1 \rangle \langle f_1|=K_n-a_nf_1$, such that $|f_1\rangle \in {\rm Ker}(A_n)$, for all $n$. Note that, in general, $A^{\dagger}_n|f_1 \rangle \ne 0$. Let us rewrite condition (\ref{C2}) in terms of $A_n$ and $A_n^{\dagger}$. It reads
\begin{equation}
\label{C2b}
\Lambda(f_1)=\sum_n(A_n+a_nf_1) f_1 (A_n^{\dagger}+a^*_nf_1)=f_1 \Rightarrow \sum_n |a_n|^2 f_1=f_1 \Rightarrow  \sum_n |a_n|^2=1,
\end{equation}
where we employed $f_1=f_1^{\dagger}$ and $A_nf_1=f_1A^{\dagger}_n=0$. Using the previous result, condition (\ref{C1}) becomes:
\begin{equation}
\label{C1b}
f_1+\sum_n a^{*}_n f_1 A_n+\sum_n a_n  A_n^{\dagger} f_1+\sum_n A^{\dagger}_n A_n =\mathds{1}. 
\end{equation}

Let us initially consider the grand sum of pure states $|\varphi \rangle$ subjected to arbitrary free operations. For any pure state, one can always write the decomposition $|\varphi \rangle= \zeta |f_1\rangle+\zeta_{\perp}|g_{\perp}\rangle$, where $\langle f_1|g_{\perp}\rangle=0$ and $|\zeta|^2+|\zeta_{\perp}|^2=1$. The grand sum of this state is $\Sigma(\varphi)=D\langle f_1|\varphi\rangle \langle \varphi |f_1\rangle=D|\zeta|^2$.
After the state is acted upon by the free channel, we get
\begin{equation}
\Lambda(\varphi)=|\zeta|^2 f_1+|\zeta_{\perp}|^2\sum_nA_n|g_{\perp}\rangle \langle g_{\perp}|A_n^{\dagger}+\zeta \zeta^{*}_{\perp}\sum_na_n|f_1\rangle \langle g_{\perp}|A_n^{\dagger}+\zeta^{*} \zeta_{\perp}\sum_na_n^{*}A_n|g_{\perp}\rangle \langle f_1|
\end{equation}
The corresponding grand sum can be written as
\begin{equation}
\label{pulo}
\Sigma(\Lambda(\varphi))=D \langle f_1|\Lambda(\varphi)|f_1\rangle=D|\zeta|^2+|\zeta_{\perp}|^2\sum_n|\langle f_1|A_n|g_{\perp}\rangle|^2+
\left[\zeta \zeta^{*}_{\perp}\sum_na_n \langle g_{\perp}|A_n^{\dagger}|f_1\rangle+ {\rm c.c.}\right],
\end{equation}
where ``c.c.'' stands for ``complex conjugate.''

Next we project condition (\ref{C1b}) on the left (right) onto $| g_{\perp} \rangle$ ($|j\rangle$):
\begin{equation}
\langle g_{\perp}|f_1|j\rangle+\sum_n a^{*}_n \langle g_{\perp}|f_1 A_n|j\rangle+\sum_n a_n \langle g_{\perp}| A_n^{\dagger} f_1|j\rangle+\sum_n \langle g_{\perp}|A^{\dagger}_n A_n|j\rangle =\langle g_{\perp}|j\rangle.
\end{equation}
Due to the orthogonality between $|f_1\rangle$ and $|g_{\perp}\rangle$, the first two terms above vanish. Using $f_1|j\rangle=\frac{1}{\sqrt{D}}|f_1\rangle$ we obtain
\begin{equation}
\frac{1}{\sqrt{D}}\sum_n a_n \langle g_{\perp}| A_n^{\dagger} |f_1\rangle+\sum_n \langle g_{\perp}|A^{\dagger}_n A_n|j\rangle =\langle g_{\perp}|j\rangle.
\end{equation}
Next, by summing over all $j$ and noticing that the first term above does not depend on this index, we get
\begin{equation}
{\sqrt{D}}\sum_n a_n \langle g_{\perp}| A_n^{\dagger} |f_1\rangle+{\sqrt{D}} \sum_n \langle g_{\perp}|A^{\dagger}_n A_n|f_1\rangle ={\sqrt{D}}\langle g_{\perp}|f_1\rangle.
\end{equation}
Since $A_n|f_1\rangle=0$ and, again, $\langle g_{\perp}|f_1\rangle=0$, we obtain
\begin{equation}
\sum_n a_n \langle g_{\perp}| A_n^{\dagger} |f_1\rangle=0.
\end{equation}
But this and its complex conjugated are proportional to the terms inside the square brackets in Eq. (\ref{pulo}). So we can finally write
\begin{equation}
\label{finalpure}
\Sigma(\Lambda(\varphi))=\Sigma(\varphi)+|\zeta_{\perp}|^2\sum_n|\langle f_1|A_n|g_{\perp}\rangle|^2 \ge \Sigma(\varphi).
\end{equation}
Therefore, we have shown that the grand sum is non-decreasing under arbitrary free operations, for pure states. Since the rugosity $\mathfrak{R}$ is a decreasing function of $\Sigma$, it does not increase under free operations, for pure states. 

It is simple to extend this result to arbitrary states. Again, we use the fact that $\varrho$ can always be written as $\varrho=\sum_{k}p_{k}\varrho^{(k)}=\sum_{k,\ell}p_{k}q^{(k)}_{\ell}|u_{\ell}^{(k)} \rangle \langle u_{\ell}^{(k)}|$, where $\{|u_{\ell}^{(k)} \rangle \}$ is the spectral basis of $\varrho^{(k)}$, with corresponding eigenvalues $q^{(k)}_{\ell}$. Thus, $\Lambda(\varrho)=\sum_{k,\ell}p_{k}q^{(k)}_{\ell}\Lambda(|u_{\ell}^{(k)} \rangle \langle u_{\ell}^{(k)}|)$ and $\Sigma(\Lambda(\varrho))=\sum_{k,\ell}p_{k}q^{(k)}_{\ell}\Sigma(\Lambda(|u_{\ell}^{(k)} \rangle \langle u_{\ell}^{(k)}|))\ge \sum_{k,\ell}p_{k}q^{(k)}_{\ell}\Sigma(|u_{\ell}^{(k)} \rangle \langle u_{\ell}^{(k)}|)=\Sigma(\varrho)$. The inequality is due to (\ref{finalpure}). Therefore, $\Sigma(\Lambda(\varrho)) \ge \Sigma(\varrho)$ and 
\begin{equation}
\mathfrak{R}(\Lambda(\varrho)) \le \mathfrak{R}(\varrho),
\end{equation} 
for arbitrary free maps and states. This is the last requirement needed to conclude that $\mathfrak{R}$ is a trusted texture measure.

\section{State Convertibility and maximum-texture states}
Now we introduce relevant free operations to establish the maximum-resource states. Consider the map $\Lambda$ defined by the following  $D^2$ Kraus 
operators
\begin{equation}
\label{kraus}
K_{n, \ell}=\frac{1}{D}\left[f_1 + \frac{\omega^n}{\sqrt{D}}\sum_{i,j=1}^D (\omega^*)^{i-1}\alpha_{i\oplus (\ell-1)}|i\oplus (\ell-1)\rangle( \langle i|-\langle j|) \right],
\end{equation}
where $|f_1\rangle=\frac{1}{\sqrt{D}}\sum_i |i\rangle$ is the only Fourier state with vanishing texture, $\oplus$ denotes sum modulo $D$, $n=1, \cdots, D$, and $\ell=1, \cdots, D$. The motivation behind this definition lies in the fact that $( \langle i|-\langle j|)$ ``detects'' texture, vanishing whenever there exists permutation symmetry between the components related to $|i\rangle$ and $|j\rangle$. 

Let us demonstrate that the map $\Lambda$ is indeed a free operation. We start by showing that $\Lambda(f_1)=f_1$. Note that $( \langle i|-\langle j|)|f_1\rangle=0$, $K_{n, \ell}|f_1\rangle=(1/D)|f_1\rangle$. It is then immediate that $K_{n,\ell}f_1 K_{n,\ell}^{\dagger}=f_1/D^2$, and $\Lambda(f_1)=\sum_{n,\ell=1}^DK_{n,\ell}f_1 K_{n,\ell}^{\dagger}=f_1$.

Next, we show that $\sum_{n,\ell=1}^DK_{n, \ell}^{\dagger}K_{n, \ell}=\mathds{1}$. The terms inside the sum are
\begin{eqnarray}
\nonumber
K_{n, \ell}^{\dagger}K_{n, \ell}=\frac{1}{D^2}\left[  |f_1\rangle \langle f_1| + \frac{(\omega^*)^n}{\sqrt{D}}\sum_{i,j=1}^D \omega^{i-1}\alpha^*_{i\oplus (\ell-1)}( | i\rangle-| j\rangle) \langle i\oplus (\ell-1)| \right]\\
\times \left[ |f_1\rangle \langle f_1| + \frac{\omega^n}{\sqrt{D}}\sum_{i,j=1}^D (\omega^*)^{i-1}\alpha_{i\oplus (\ell-1)}|i\oplus (\ell-1)\rangle( \langle i|-\langle j|) \right].
\end{eqnarray}
Out of the four resulting terms, two depend on $n$ exclusively through the multiplicative factor $\omega^n$. When the sum in $n$ is effected, because $\sum_{n=1}^D\omega^n=\sum_{n=1}^D(\omega^*)^n=0$, the corresponding terms vanish. Therefore
\begin{eqnarray}
\nonumber
\sum_{n=1}^DK_{n, \ell}^{\dagger}K_{n, \ell}&=&\frac{1}{D^2}\left[ D |f_1\rangle \langle f_1| + \sum_{i,j=1}^D\sum_{r,s=1}^D \omega^{i-1}\alpha^*_{i\oplus (\ell-1)} (\omega^*)^{r-1}\alpha_{r\oplus (\ell-1)}( | i\rangle-| j\rangle)( \langle r|-\langle s|)\delta_{i,r}\right]\\
&=&\frac{1}{D^2}\left[ D |f_1\rangle \langle f_1| +\sum_{i,j,s=1}^D|\alpha_{i\oplus (\ell-1)}|^2( | i\rangle-| j\rangle)( \langle i|-\langle s|)\right].
\end{eqnarray}
By summing over $\ell$ we get
\begin{eqnarray}
\nonumber
\sum_{n,\ell=1}^DK_{n, \ell}^{\dagger}K_{n, \ell}&=&\frac{1}{D^2}\left[ D^2 |f_1\rangle \langle f_1| + \sum_{i,j,s=1}^D\left(\sum_{\ell=1}^D|\alpha_{i\oplus (\ell-1)}|^2\right)( | i\rangle-| j\rangle)( \langle i|-\langle s|)\right]\\
\nonumber
&=&|f_1\rangle \langle f_1| + \frac{1}{D^2}\sum_{i,j,s=1}^D( | i\rangle-| j\rangle)( \langle i|-\langle s|)=|f_1\rangle \langle f_1| + \frac{1}{D^2}\left(D^2\sum_{i=1}^D | i\rangle \langle i| -D\sum_{i,j=1}^D | i\rangle \langle j| \right)\\
&=&|f_1\rangle \langle f_1| + \mathds{1}  -|f_1\rangle \langle f_1|=\mathds{1},
\end{eqnarray}
where we used $\sum_{\ell=1}^D|\alpha_{i\oplus (\ell-1)}|^2=1$ and $D|f_1\rangle \langle f_1|=\sum_{i,j=1}^D | i\rangle \langle j| $.

Now we address the maximal states. For definiteness, out of the Fourier states, we take $|f_2\rangle$ as the maximum-resource state from which one can deterministically produce any pure state $|\psi\rangle=\sum_i\alpha_i |i\rangle$ (the case of mixed states will be addressed later). Of course, we could have picked any $| f_k \rangle$ with $k>1$, with simple adjustments in the Kraus operators. 
Note also that summation in the index $j$, in Eq. (\ref{kraus}) leads to
\begin{equation}
K_{n, \ell}=\frac{1}{D}\left[ |f_1\rangle \langle f_1| + \frac{\omega^n}{\sqrt{D}}\sum_{i=1}^D (\omega^*)^{i-1}\alpha_{i\oplus (\ell-1)}|i\oplus (\ell-1)\rangle(D \langle i|-\sqrt{D}\langle f_1|) \right].
\end{equation}
This last form makes it clear that
\begin{equation}
K_{n, \ell}|f_2\rangle=\frac{\omega^n}{\sqrt{D}}\sum_{i=1}^D (\omega^*)^{i-1}\alpha_{i\oplus (\ell-1)}|i\oplus (\ell-1)\rangle \langle i|f_2\rangle,
\end{equation}
because $\langle f_k|f_j\rangle=\delta_{kj}$. Since $\langle i|f_2\rangle=\omega^{i-1}/\sqrt{D}$, we get
\begin{equation}
K_{n, \ell}|f_2\rangle=\frac{\omega^n}{D}\sum_{i=1}^D\alpha_{i\oplus (\ell-1)}|i\oplus (\ell-1)\rangle=\frac{\omega^n}{D}|\psi\rangle.
\end{equation}
Therefore the map produces $|\psi\rangle$ deterministically: 
\begin{equation}
K_{n, \ell}|f_2\rangle \langle f_2|K_{n, \ell}^{\dagger}=\frac{1}{D^2}|\psi\rangle \langle \psi |\;\; \forall \;\; n, \ell.
\end{equation}
Finally,  we show that an arbitrary mixed state can also be produced from $|f_2\rangle$ and free operations. To do this we parallel the arguments in the Supplemental  material of \cite{l1}. Consider an arbitrary mixed state given by $\rho=\sum_{k}q_k|\psi_k\rangle \langle \psi_k|$, where $|\psi_k\rangle=\sum_i\alpha^{(k)}_i|i\rangle$ and $\sum_k q_k=1$, $0\le q_k \le 1$. From the previous results, it is evident that the Kraus operators
\begin{equation}
K_{n, \ell,k}=\frac{\sqrt{q_k}}{D}\left[ |f_1\rangle \langle f_1| + \frac{\omega^n}{\sqrt{D}}\sum_{i,j=1}^D (\omega^*)^{i-1}\alpha^{(k)}_{i\oplus (\ell-1)}|i\oplus (\ell-1)\rangle( \langle i|-\langle j|) \right],
\end{equation}
represent free operations and sum to unity. Also from the previous results we see that $K_{n, \ell,k}|f_2\rangle \propto |\psi_k\rangle$  $\forall \; n, \ell$. The non-selective POVM effected by these operators over the ket $|f_2 \rangle$ generates the state:
\begin{equation}
\rho=\sum_{n, \ell, k}K_{n, \ell,k}|f_2 \rangle \langle f_2| K_{n, \ell,k}^{\dagger}=\sum_k q_k |\psi_k\rangle \langle \psi_k|,
\end{equation}
which is the desired result.
\newpage

\section{Out-of-equilibrium paramagnet }
Let us illustrate how quantum state texture arises in a simple non-equilibrium scenario, and how it relates to equilibrium thermodynamic quantities. 
Consider a general system in thermal equilibrium with a reservoir at absolute temperature $T$.
The system is appropriately described by the canonical Gibbs state:
\begin{equation}
\label{thermal}
 \rho_G(T)=\frac{1}{\cal Z}\sum_i e^{- E_i/k_B T}|i\rangle \langle i|,
 \end{equation}
where ${\cal Z}$ is the canonical partition function and $k_B$ is the Boltzmann constant. To a large extent, thermodynamics is about fundamental constraints on energy conversion. It is, therefore, natural to investigate quantum texture in the energy basis, which we take as $\{|i\rangle\}$. Since, in this basis, the state (\ref{thermal}) has no coherences, we simply obtain $\mathfrak{R}(\rho_G)=\ln D$, independently of $T$. In this specific context, a variation in quantum texture as the temperature changes is an indicator of a non-equilibrium situation.

One interesting class of out-of-equilibrium states is that of coherent Gibbs kets, 
given by 
\begin{equation}
\label{gibbs1}
|\Psi_G\rangle=\frac{1}{\sqrt{\cal Z}}\sum_i e^{- E_i/2k_B T}|i\rangle,
\end{equation}
which have the desirable property of becoming the equilibrium state (\ref{thermal}) under full decoherence. However, there is no physical reason for the coherences of this state to be all real and positive. A less contrived definition with the same equilibration property is:
\begin{equation}
\label{gibbs2}
|\Psi_{\{\phi_j\}}\rangle=\frac{1}{\sqrt{\cal Z}}\sum_j e^{i \phi_j}e^{- E_j/2k_B T}|j\rangle,
\end{equation}
where the phases $\{\phi_j\}$ are uniformly and independently distributed over the interval $[0,2\pi)$. Using state (\ref{gibbs1}) rather than (\ref{gibbs2}) is to completely suppress the imaginarity of quantum states. Note that the $\ell_1$-norm of coherence is the same for these two states, while the rugosity is not. 

Consider the ideal paramagnet, a non-interacting spin chain, with Hamiltonian $H=-\mu_0B\sum_j\sigma_j$, where $\mu_0$ is the magnetic moment and $B$ is the external magnetic field. Each spin ($D=2$) can assume two energies, $E_1=-\mu_0B$ (up) and $E_2=+\mu_0B$ (down), since $\sigma_j=\pm1$. The canonical partition function per spin is ${\cal Z}= 2 \cosh(\mu_0B/k_BT)$. Regarding the random phases, for simplicity, one can set $\phi_1=0$ and $\phi_2\equiv \phi$. The grand sum of $\Psi_{\phi}$ reads 
\begin{equation}
\Sigma(\Psi_{\phi})=1+ \sech\left(\frac{\mu_0B}{k_BT}\right) \cos \phi,
\end{equation}
with $\mathfrak{R}(\Psi_{\phi})=-\ln [\Sigma(\Psi_{\phi})/2]$. 

Now consider a large number of such systems, in contact with the same reservoir, and, thus, differing only by their random phases.
The additivity property implies 
\begin{equation}
\mathfrak{R}(\Psi_{\phi_{1}}\otimes \cdots \otimes \Psi_{\phi_{N}})=\sum_{\nu=1}^N\mathfrak{R}(\Psi_{\phi_{\nu}})\equiv \mathfrak{R}(T),
\end{equation}
where $N$ is the number of spins. In approaching the thermodynamic limit ($N\rightarrow \infty$) one can replace the sum with the integral:
\begin{equation}
\frac{\mathfrak{R}(T)}{N}=\int_0^{2\pi}\mathfrak{R}(\Psi_{\phi}) \mu(\phi)d\phi=-\frac{1}{2\pi}\int_0^{2\pi} \ln \frac{\Sigma(\Psi_{\phi})}{2}d\phi,
\end{equation}
where we used the uniform distribution $\mu(\phi)=1/2\pi$. The solution of this integral is \cite{grad}:
\begin{equation}
\mathfrak{R}(T)=-N\ln\left(\frac{1}{2}+\frac{m(T)}{2N\mu_0} \right),
\end{equation}
where $m(T)=N\mu_0\tanh \left(\frac{\mu_0B}{2k_BT}\right)$ is the magnetization of the chain when it is in thermal equilibrium at temperature $T$. Therefore, the rugosity of the non-equilibrium system, in the thermodynamic limit, can be expressed in terms of the magnetization of the corresponding equilibrated state.

\newpage
\section{Circuit layer identification protocol (detailed)}
Let us start by considering arbitrary input qubits, parametrized as $|\psi_{\rm in}\rangle=\cos(\theta/2)|+\rangle+e^{i\phi}\sin(\theta/2)|-\rangle$. Of course, 
\begin{equation}
\Sigma(\psi_{\rm in})\equiv\Sigma_{\rm in}=1+2\cos(\theta/2)\sin(\theta/2)\cos \phi.
\end{equation}
Now we average over $\theta$ and $\phi$ using the Haar measure:
\begin{equation}
\overline{\Sigma}_{\rm in}=\frac{1}{4\pi}\int_0^{2\pi} \int_0^{\pi}\Sigma_{\rm in}(\theta,\phi)\sin\theta \, d \theta d\phi=\frac{1}{4\pi}\int_0^{2\pi} \int_0^{\pi}\sin \theta\, d \theta d\phi=1,
\end{equation}
because $\int_0^{2\pi}\cos \phi \, d \phi=0$. Note that one can alternatively write 
\begin{equation}
\overline{\Sigma}_{\rm in}=\int d\Omega \, D{\rm Tr}\left( f_1 \psi_{\rm in}  \right)= 2 {\rm Tr}\left( f_1\int d\Omega \, \psi_{\rm in}\right)=1,
\end{equation}
where $D=2$, $4\pi d\Omega=\sin \theta\, d \theta d\phi$, and $ \psi_{\rm in}=|\psi_{\rm in}\rangle \langle \psi_{\rm in}|$. We also used $\int d\Omega \, \psi_{\rm in}=\frac{\mathds{1}}{2}$.
We now show that this average remains unchanged after the action of an arbitrary single-qubit unitary $U$. In this case, the output average reads
\begin{eqnarray}
\nonumber
\overline{\Sigma}_{\rm out}&=&\int d\Omega \, D{\rm Tr}\left(f_1  U \psi_{\rm in}U^{-1} \right)= 2 {\rm Tr}\left( f_1\int d\Omega \, U\psi_{\rm in}U^{-1}\right)=2 {\rm Tr}\left( f_1U\left(\int d\Omega \, \psi_{\rm in}\right)U^{-1}\right)\\
&=&2 {\rm Tr}\left( f_1U\frac{\mathds{1}}{2}U^{-1}\right)= {\rm Tr}( f_1)=1
\end{eqnarray}
Therefore, we get $\overline{\Sigma}_{\rm out}=1$. In particular
\begin{eqnarray}
\overline{\Sigma}_{\rm out}^{\rm T}=\overline{\Sigma}_{\rm out}^{\rm S}=\overline{\Sigma}_{\rm out}^{\rm H}=1=\overline{\Sigma}_{\rm in},
\end{eqnarray}
which is the result described for the single-qubit gates in Eq. (4) of the main text.

We are left with the grand sum related to the qubits that go through CNOT gates. In this case, we have the following input two-qubit state:
\begin{equation}
|\Psi_{\rm in}\rangle=|\psi_{\rm in}\rangle \otimes |\psi_{\rm in}\rangle=a^2|++\rangle+ab|+-\rangle+ba|-+\rangle+b^2|--\rangle,
\end{equation}
where we defined $a=\cos(\theta/2)$ and $b=e^{i\phi}\sin(\theta/2)$. Note that, due to the fact that the two qubits have the same input states in a given run, this qudit ($D=4$) state does not satisfy the Haar measure in the larger Hilbert space.
This is instrumental for the understanding of the results that follow. After the action of a CNOT gate the output state reads
\begin{eqnarray}
|\Psi_{\rm out}\rangle=a^2|++\rangle+ab|+-\rangle+b^2|-+\rangle+ba|--\rangle
\end{eqnarray}
Since, at the end of the circuit layer we measure individual qubits, we must determine the reduced density matrices for both, the control and the target qubits.
After some simple algebra we obtain
\begin{eqnarray}
\varrho_{\rm out}^{\mbox{\large \textbullet}}&=&|a|^2|+\rangle \langle+| + (a^2{b^*}^2+|a|^2|b|^2)|+\rangle \langle-|+ ({a^*}^2{b^*}+|a|^2|b|^2)|-\rangle \langle+| +
|b|^2|-\rangle \langle -|, 
\end{eqnarray}
\begin{eqnarray}
\varrho_{\rm out}^{\oplus}&=&
(|a|^4+|b|^4)|+\rangle \langle+| + (|a|^2ab^*+|b|^2a^*b)|+\rangle \langle-|+  (|a|^2a^*b+|b|^2ab^*)|-\rangle \langle+| +
2|a|^2|b|^2|-\rangle \langle -|, 
\end{eqnarray}
where ``{\large \textbullet}'' (``$\oplus$'') refers to control (target) qubit, i. e., $\varrho_{\rm out}^{\mbox{\large \textbullet}}=\Tr_{\oplus}(\Psi_{\rm out})$ and  $\varrho_{\rm out}^{\oplus}=\Tr_{\mbox{\large \textbullet}}(\Psi_{\rm out})$. Finally one can calculate the associated grand sums, in the known basis, which read
\begin{eqnarray}
\nonumber
\Sigma_{\rm out}^{\mbox{\large \textbullet}}&=&1+\langle 1|\varrho_{\rm out}^{\mbox{\large \textbullet}}|2\rangle +\langle 2|\varrho_{\rm out}^{\mbox{\large \textbullet}}|1\rangle \\
&=&1+(\beta^2+{\beta^*}^2-\alpha^2-{\alpha^*}^2)|a|^2|b|^2+(\alpha \beta^*+\alpha \beta^*)(|a|^2-|b|^2)+[(\beta^2-\alpha^2)a^2{b^*}^2+\mbox{c.c}],
\end{eqnarray}
and
\begin{eqnarray}
\nonumber
\Sigma_{\rm out}^{\oplus}&=&1+\langle 1|\varrho_{\rm out}^{\oplus}|2\rangle +\langle 2|\varrho_{\rm out}^{\oplus}|1\rangle \\
&=&1+(\alpha \beta^*+\alpha^* \beta)(1-4|a|^2|b|^2)+[(\beta^2-\alpha^2)(|a|^2ab^*+|b|^2a^* b)+\mbox{c.c}],
\end{eqnarray}
where we used $|a|^4-|b|^4=|a|^2-|b|^2$; and $\langle 1|+\rangle=\alpha$,  $\langle 1|-\rangle=\beta^*$,  $\langle 2|+\rangle=\beta$, and  $\langle 2|-\rangle=-\alpha^*$. The terms $\sim (|a|^2-|b|^2)$ have a vanishing average:
\begin{equation}
\overline{|a|^2}=\frac{1}{4\pi}\int_0^{2\pi}d\phi \int_0^{\pi}\cos^2(\theta/2)\sin(\theta)d \theta=\frac{1}{4\pi}\int_0^{2\pi}d\phi \int_0^{\pi}\sin^2(\theta/2)\sin(\theta)d \theta=\overline{|b|^2}=\frac{1}{2}.
\end{equation}
The same happens for the terms $\sim b $ and $\sim b^*$, because $\int_0^{2\pi}d\phi e^{i\phi}=0$.
Finally, the terms $\sim b^2 $ and $\sim {b^*}^2$ also vanish, because $\int_0^{2\pi}d\phi e^{2i\phi}=0$. In addition
\begin{equation}
\overline{|a|^2|b|^2}=\frac{1}{4\pi}\int_0^{2\pi}d\phi \int_0^{\pi}\cos^2(\theta/2)\sin^2(\theta/2)\sin(\theta)d \theta=\frac{1}{6},
\end{equation}
so, when averaging over the random inputs, we obtain:
\begin{eqnarray}
\label{cnot1a}
\overline{\Sigma}_{\rm out}^{\mbox{\large \textbullet}}&=&1-\frac{1}{6}(\alpha^2+{\alpha^*}^2-\beta^2-{\beta^*}^2),\\
\label{cnot1b}
\overline{\Sigma}_{\rm out}^{\oplus}&=&1+\frac{1}{3}({\alpha^*}\beta+\alpha{\beta^*}),
\end{eqnarray}
which together with [$\alpha \rightarrow (\alpha+\beta)/\sqrt{2}$ and $\beta \rightarrow (\alpha-\beta)/\sqrt{2}$]
\begin{eqnarray}
\label{cnot2a}
\overline{\Sigma'}_{\rm out}^{\mbox{\large \textbullet}}&=&1-\frac{1}{3}({\alpha}\beta+{\alpha^*}{\beta^*}),\\
\label{cnot2b}
\overline{\Sigma'}_{\rm out}^{\oplus}&=&1+\frac{1}{3}(|\alpha|^2-|\beta|^2),
\end{eqnarray}
correspond to the results described in the main text. 

We now demonstrate inequality (10) of the main text. We start from
\begin{equation}
\left({\rm Re}(\alpha^2)+{\rm Re}(\beta^2)\right)^2\ge 0 \Rightarrow (\alpha^2+{\alpha^*}^2+\beta^2+{\beta^*}^2)^2\ge 0.
\end{equation}
But
\begin{eqnarray}
\nonumber
(\alpha^2+{\alpha^*}^2+\beta^2+{\beta^*}^2)^2=(\alpha^2+{\alpha^*}^2-\beta^2-{\beta^*}^2)^2+4(\alpha^2+{\alpha^*}^2)(\beta^2+{\beta^*}^2)\\
\nonumber
=36\left(\overline{\Sigma}_{\rm out}^{\mbox{\large \textbullet}}-1\right)^2+4[({\alpha^*}\beta+\alpha{\beta^*})^2+({\alpha}\beta+{\alpha^*}{\beta^*})^2-4|\alpha|^2|\beta|^2]\\
=\left(\overline{\Sigma}_{\rm out}^{\mbox{\large \textbullet}}-1\right)^2+\left(\overline{\Sigma}_{\rm out}^{\oplus}-1\right)^2+\left(\overline{\Sigma'}_{\rm out}^{\mbox{\large \textbullet}}-1\right)^2
+\left(\overline{\Sigma'}_{\rm out}^{\oplus}-1\right)^2-\frac{1}{9},
\end{eqnarray}
which corresponds to the final result. It is evident that $\overline{\Sigma}_{\rm out}^{\mbox{\large \textbullet}}=1$,  $\overline{\Sigma}_{\rm out}^{\oplus}=1$, $\overline{\Sigma'}_{\rm out}^{\mbox{\large \textbullet}}=1$, and $\overline{\Sigma'}_{\rm out}^{\oplus}=1$, simultaneously is impossible. Suppose instead that three of these quantities equal 1. This can only happen by fine-tuning the parameters, e. g., $\alpha=1/\sqrt{2}$ and $\beta=i/\sqrt{2}$. This leads to $\overline{\Sigma}_{\rm out}^{\oplus}=\overline{\Sigma'}_{\rm out}^{\mbox{\large \textbullet}}=\overline{\Sigma'}_{\rm out}^{\oplus}=1$, but $\overline{\Sigma}_{\rm out}^{\mbox{\large \textbullet}}=2/3$, which enables the determination of CNOT gates. In addition, such a result means that the unknown basis is a balanced superposition of the elements of the known basis with a relative phase of $\pm \pi/2$. Therefore, simple tests (see below) would allow for the determination of the $|\pm\rangle$ basis.

In general, the determination of $\alpha=|\alpha| e^{i\lambda}$ and $\beta=|\beta| e^{i\chi}$ in terms of the measured four numbers is as follows. Let us initially consider 
\begin{equation}
\label{notation}
X=\overline{\Sigma}_{\rm out}^{\mbox{\large \textbullet}}, \;\; \tilde{X}=\overline{\Sigma}_{\rm out}^{\oplus},  \;\; \tilde{Y}=\overline{\Sigma'}_{\rm out}^{\oplus},\;\; \mbox{and} \;\; \tilde{Y}=\overline{\Sigma'}_{\rm out}^{\oplus}.
\end{equation}
From $\tilde{Y}=1+\frac{1}{3}(|\alpha|^2-|\beta|^2)$, Eq. (\ref{cnot2b}) and $|\alpha|^2+|\beta|^2=1$ one can immediately write
\begin{equation}
|\alpha|=\sqrt{\frac{3}{2}\tilde{Y}-1},\;\; |\beta|=\sqrt{2-\frac{3}{2}\tilde{Y}}.
\end{equation}
Equation (\ref{cnot1a}) can be rewritten as
\begin{equation}
\frac{3}{2}(X-\tilde{Y})=-|\alpha|^2\cos^2(\lambda)+|\beta|^2\cos^2(\chi).
\end{equation}
Combining Eqs. (\ref{cnot1b}) and (\ref{cnot2a}):
\begin{equation}
\frac{3}{2}(\tilde{X}-Y)=-|\alpha||\beta|\cos(\lambda)\cos(\chi).
\end{equation}
From these two last equations we get
\begin{eqnarray}
\label{final}
\cos \lambda&=& \pm\frac{\sqrt{\tilde{Y}-X+\sqrt{(\tilde{Y}-X)^2+(\tilde{X}-Y)^2}}}{\sqrt{2\tilde{Y}-\frac{4}{3}}},\\
\cos \chi&=& \pm\frac{\sqrt{X-\tilde{Y}+\sqrt{(\tilde{Y}-X)^2+(\tilde{X}-Y)^2}}}{\sqrt{\frac{8}{3}-2\tilde{Y}}},
\end{eqnarray}
which finishes the demonstration. 

Alternatively, by  adopting only the positive signs above, one have to check the following bases: $|+\rangle=\alpha|1\rangle+\beta|2\rangle$, $|+\rangle=\alpha|1\rangle-\beta|2\rangle$, $|+\rangle=-\alpha|1\rangle+\beta|2\rangle$, and $|+\rangle=-(\alpha|1\rangle+\beta|2\rangle)$, the last two being redundant (to get the corresponding expressions for $|-\rangle$ we just impose orthogonality). We thus have two candidate bases to test.
So we reduced the possible unknown bases from infinitely many to two. From this it is easy to decide which is the correct basis. 

Assume that $|+\rangle=\alpha|1\rangle+\beta|2\rangle$ and $|-\rangle=\beta^*|1\rangle-\alpha^*|2\rangle$ give the correct relation between the known and unknown bases. Since we know where the CNOT gates are we just have to feed $|+\rangle=\alpha|1\rangle+\beta|2\rangle$ as the input qubits of all tracks leading to CNOT gates. If the output is $|+\rangle$ for all output qubits, then the initial hypothesis is true. 
Otherwise, redefine the ``$+$'' as $|+\rangle=\alpha|1\rangle-\beta|2\rangle$ and send through the CNOT gates. If the output is $|+\rangle$ for all output qubits, then the second hypothesis is true.
If the procedure fails in both cases then, the attributions in (\ref{notation}) are wrong. We just have to make $\tilde{X}\leftrightarrow X$ and $\tilde{Y}\leftrightarrow Y$ and repeat the process for the new $\alpha$ and $\beta$. One of the four bases, the correct one, will output only ``$+$'', which finishes the characterization. In particular this would allow for the identification of all single qubit gates.

\end{widetext}

\end{document}